\documentclass[seceq,preprint]{ptptex}

\def\nn{\nonumber\\}

\def\ep{{\epsilon}}

\newcommand{\bra}[1]{\left\langle{#1}\right|}
\newcommand{\ket}[1]{\left|{#1}\right\rangle}

\renewcommand{\bar}[1]{\overline{#1}}

\def\d{d}

\def\del{{\partial}}

\def\omm{\omega^{(-)}}

\def\dal{\Box}

\def\Im{\textrm{Im}\ \!}
\def\Re{\textrm{Re}\ \!}

\notypesetlogo                       
\preprintnumber[3cm]{
YITP-09-06\\ February 2009}

\title{
Supersymmetric $AdS_3$ solutions\\ in Heterotic Supergravity%
}

\author{
Hiroshi {\scshape Kunitomo}\footnote{E-mail: kunitomo@yukawa.kyoto-u.ac.jp} 
and Mitsuhisa {\scshape Ohta}\footnote{E-mail: mituhisa@yukawa.kyoto-u.ac.jp}%
}

\inst{
Yukawa Institute for Theoretical Physics, Kyoto University,\\
Kyoto 606-8502, Japan
}

\recdate{May 14, 2009}

\abst{
We analyze the $AdS_{3} \times \mathcal{M}_{7}$-type supersymmetric solutions,
including nontrivial fluxes, of Killing spinor equations in heterotic supergravity. 
We classify these solutions using their $G$-structures and intrinsic torsions,
for the cases that the number of seven-dimensional Killing spinors $N$ are equal to $1,2,3$ and $4$. 
We find that the solutions cannot have a nontrivial warp factor and
the seven-dimensional manifold $\mathcal{M}_{7}$ is characterized by 
$G_2$($SU(3)$)-structures for the $N=1$ $(2)$ case and
an $SU(2)$-structure for the $N=3$ and $4$ cases. They are further classified using their nontrivial intrinsic torsions.
It is shown,  including the leading-order $\alpha'$-corrections,
that if we impose the Bianchi identities, 
the integrability conditions of the Killing spinor equations imply
that all the field equations are satisfied.
}

\begin{document}

\maketitle

\section{Introduction}\label{sec:intro}

It is important to investigate the classical solutions of supergravity
that preserve some supersymmetries. In particular, solutions including $AdS$-space 
are interesting from the viewpoint of the $AdS/CFT$ correspondence \cite{Maldacena:1997re}.
They provide us with a deeper understanding of the dynamics of both gauge theory and gravity.

In particular, the $AdS_{3}/CFT_{2}$ correspondence in heterotic string theory has recently been studied 
\cite{Dabholkar:2007gp, Lapan:2007jx, Kraus:2007vu, Hohenegger:2008du},
in which it was conjectured to exist as a $CFT$ dual of a geometry describing 
a fundamental heterotic string. 
A mini-black-string solution \cite{Castro:2007hc, Dabholkar:2004dq} was considerd
in the five-dimensional heterotic supergravity, obtained by $T^5$-compactification, 
with $R^{2}$ correction terms \cite{Hanaki:2006pj}.
Without $R^2$ corrections, the solution has a zero horizon area
and thus there is no room for the $AdS$-space to appear.
However, once the corrections are included, the horizon is stretched \cite{Sen:2005kj}
and its near-horizon geometry becomes the $AdS_3$-space.
Unfortunately, however, such a mini-black-string solution is yet unknown 
in the ten-dimensional heterotic supergravity, which is required for more general compactifications.
It is interesting, therefore, to study supersymmetric classical solutions 
in the form of $AdS_{3}\times \mathcal{M}_{7}$, admitting a warp factor in general, 
taking into account the $R^2$ corrections.

In general, manifolds described by supersymmetric classical solutions with nontrivial fluxes
are characterized by $G$-structure, which are generalizations of 
those with special holonomy\cite{Gauntlett:2005bn}. 
As is well known, if we set all the fluxes to zero, supersymmetric solutions 
yield some special holonomy manifolds. However, if the fluxes are switched on,
they no longer have special holonomy but are characterized by $G$-structure.
The main difference between the two characterizations is on the differential conditions 
on the characteristic forms.
For example, let us consider a three fold with $SU(3)$ holonomy, a Calabi-Yau (CY) manifold.
The characteristic forms in the CY manifold are two closed forms, 
namely a K\"ahler form and a holomorphic three-form.
If we relax the closedness conditions for these forms, the three-fold no longer has 
$SU(3)$ holonomy but is characterized by more general $SU(3)$-structure.\footnote{
For the $SU(3)$-structure manifold, these forms are called $SU(3)$-invariant forms.
} 
The deviation away from the special-holonomy manifold is measured using the intrinsic torsion
of the $SU(3)$-structure.
In summary, the geometry of the supersymmetric solutions with fluxes is characterized by 
$G$-structure, which is further classified by its intrinsic torsion.\footnote{Classification and 
analysis of the supersymmetric solutions for general compactification 
with non-trivial fluxes were studied in Refs.~\citen{Fernandez:2008wa,Fernandez:2008aa,Fernandez:2008pf} 
in heterotic string theory, type II string theory,
\cite{Gauntlett:2001ur, Gauntlett:2002sc, Gauntlett:2003cy, Gran:2005wn, 
Gauntlett:2006af,Gauntlett:2007ts, Donos:2008ug, Donos:2008hd}
and M theory \cite{Gauntlett:2007ts, Behrndt:2005im, Gauntlett:2006qw}.}

At the same time, we can show that all the equations of motion are automatically
satisfied for the  $AdS_{3}\times \mathcal{M}_{7}$-type space-time,
if we impose the Killing spinor equations and the Bianchi identities.
This holds including $R^{2}$ corrections, or equivalently the leading 
$\alpha'$ correction.
In this regard, however, the Killing spinor equations are special in the sense that 
they do not have the leading $\alpha'$-corrections \cite{Gillard:2005ic}. 
In other words, the $\alpha'$-corrections appear only through the Bianchi identities.

In this paper, we classify the solutions of Killing spinor equations, 
in the form of $AdS_{3}\times \mathcal{M}_{7}$, for the cases that the number of 
seven-dimensional Killing spinors $N$ are equal to $1,2,3$, and $4$ 
in ten-dimensional heterotic supergravity.\footnote{
For more general analysis without $AdS_{3}\times \mathcal{M}_{7}$ ansatz, see
Refs.~\citen{Gran:2005wf, Gran:2007fu, Gran:2007kh, Papadopoulos:2008rx}.} 
We found that the warp factor has to be constant, thus the space-time takes the form 
of the direct product of $AdS_{3}$ and $\mathcal{M}_{7}$. The seven-dimensional manifold
$\mathcal{M}_{7}$ admits $G_{2}$ or $SU(3)$-structure for the case of
$N=1$ or $N=2$, respectively. For the $N=3$ and $4$ cases, it is characterized by
the $SU(2)$-structure again.
The intrinsic torsions, further classifying these $G$-structure manifolds, are also given.

The paper is organized as follows.
To fix our conventions, we summarize the equations of motion, Bianchi 
identities, and the Killing spinor equations in the ten-dimensional heterotic
supergravity with the leading-order $\alpha'$-corrections in \S \ref{sec:hetero}.
In \S \ref{sec:g-structure}, we introduce a fermion representation of $SO(7)$ spinors 
and simplify general spinors by choosing a special local Lorentz frame \cite{Gillard:2004xq, Gran:2006pe}.
The bilinear forms of the Killing spinors are introduced in \S \ref{sec:killing}.
These forms are invariant under some group $G\ (\subset SO(7))$ and give $G$-invariant
forms defining the $G$-structure. They satisfy the differential equations derived from
the Killing spinor equations, which are computed in \S \ref{sec:results}.
They can be interpreted in terms of the intrinsic torsion and fall into a number of classes. 
In \S \ref{sec:integrability},
we show that Killing spinor equations and Bianchi identities 
imply all the equations of motion with the leading-order $\alpha'$-corrections.
The final \S \ref{sec:summary} is devoted to summarizing the results. 
In addition, two appendices are also given. 
In Appendix \ref{sec:conventions}, we summarize the conventions for a representation 
of the $SO(7)$ gamma matrices used in the text.
The decompositions of general fluxes into $G$-representations are given 
in Appendix \ref{sec:decomp}.

\section{Heterotic supergravity on $AdS_{3} \times \mathcal{M}_{7}$}
\label{sec:hetero}

In this section, we present the equations of motion, Bianchi identities, and Killing spinor equations 
including the leading $\alpha'$-corrections to fix our conventions.
After assuming some ansatz on metric and fluxes, we rewrite the Killing spinor equations
on $AdS_{3}\times\mathcal{M}_7$.

\subsection{Equations of motion, Bianchi dentities and Killing spinor equations}

The equations of motion in heterotic supergravity,
\cite{Sen:1985qt, Callan:1985ia, Hull:1986kz} including $\alpha'$-order corrections, are given by
\begin{eqnarray}
 E_{(MN)} &\equiv& R_{MN} + 2\nabla_{M}\nabla_{N}\Phi - \frac{1}{4}H_{MPQ}{H_{N}}^{PQ}
  \nn
  && \qquad - 2 \alpha'
  \left(
   \mathrm{tr}(F_{MP}{F_{N}}^{P})
   - \mathrm{tr}(R^{(+)}_{MP}{R^{(+)}_{N}}{}^{P})
  \right) = 0, \\
 E &\equiv& (\nabla\Phi)^{2} - \frac{1}{2}\dal\Phi
  - \frac{1}{4\cdot 3!}H_{MNP}H^{MNP} \\
 && \qquad - \frac{\alpha'}{4}
  \left(
   \mathrm{tr}(F_{MN}F^{MN}) - \mathrm{tr}(R^{(+)}_{MN}R^{(+)MN})
  \right) = 0, \\
 E_{[MN]} &\equiv& \nabla_{P}(e^{-2\Phi}{H^{P}}_{MN}) = 0, \\
 E_{M} &\equiv& \mathcal{D}^{(-)}_{N}(e^{-2\Phi}{F^N}_M) = 0.
\end{eqnarray}
In addition, the Bianchi identities for three-form flux $H$ and gauge field strength $F$ are given by
\begin{equation}
 \d H = 2\alpha'
	\left(
	 \mathrm{tr}(R^{(+)}\wedge R^{(+)})
	 - \mathrm{tr}(F\wedge F)
	\right),
\end{equation}
and
\begin{equation}
 \d F = 0,
\end{equation}
respectively.

The classical solution of these equations of motion is called supersymmetric,
if we further require the supersymmetry transformation of the fermion fields to vanish,
which imposes the preseration of the supersymmetry on the solution.

For heterotic supergravity, the conditions for preserving the supersymmetry are given by
\begin{eqnarray}
 \delta\Psi_{M}
  &=& \nabla^{(-)}_{M}\ep \equiv \nabla_{M}\ep
  - \frac{1}{4\cdot2!}H_{MNP}\Gamma^{NP}\ep = 0, \label{KSE}\\
 \delta\lambda
  &=& -\frac{1}{2}\nabla_{M}\Phi\Gamma^{M}\ep
  + \frac{1}{4\cdot3!}H_{MNP}\Gamma^{MNP}\ep = 0, \\
 \delta\chi
  &=& -\frac{1}{2\cdot2!}F_{MN}\Gamma^{MN}\ep = 0,
\end{eqnarray}
where $\Psi_M$, $\lambda$, and $\chi$ are gravitino, dilatino, and gaugino,
respectively \cite{Bergshoeff:1989de, Strominger:1986uh}.
We call them Killing spinor equations because the first one (\ref{KSE}) has the form
of the conventional Killing spinor equation if we regard $\nabla^{(-)}$ as 
a covariant derivative with torsion. It should be noted that these Killing spinor equations 
are not modified up to $\alpha'^{2}$-order corrections.

\subsection{Ansatz for metric, fluxes, and spinor}

We assume that the ten-dimensional metric takes the form of
\begin{equation}
 \d s^{2} 
  = e^{-2A(y)}\tilde{g}_{\mu\nu}(x)\d x^{\mu}\d x^{\nu} + g_{mn}(y)\d y^{m}\d y^{n},
\end{equation}
where $\mu,\nu=0,1,2$ and $ m,n=1,\cdots,7$. 
The metrics $\tilde{g}_{\mu\nu}(x)$ and $g_{mn}(y)$ are those of $AdS_3$ and $\mathcal{M}_7$
manifolds, respectively. 
The warp factor $A(y)$ is assumed to depend only on the six-dimensional coordinates $\{y^m\}$.

We also assume the forms of the three-form $H$ and the gauge field $F$ as
\begin{eqnarray}
H&=&\sqrt{-\tilde{g}(x)}e^{-3A(y)}h(y)dx^0\wedge dx^1\wedge dx^2+
\frac{1}{3!}H_{lmn}(y)dy^l\wedge dy^m\wedge dy^n,\\
F&=&\frac{1}{2}F_{mn}(y)dy^m\wedge dy^n,
\end{eqnarray} 
which are consistent with the isometry of the $AdS_3\times\mathcal{M}_7$ space-time.

By definition, there is, at least, one Killing spinor $\eta$ when we consider supersymmetric 
classical solutions. A ten-dimensional Majorana-Weyl spinor $\eta$ satisfying
\begin{eqnarray}
 \eta^{c} &=& C_{10}\bar\eta^T=\eta, \label{eq:etaismajo}\\
 \Gamma^{\natural}\eta &=& \eta, \label{eq:etaisweyl}
\end{eqnarray}
can be decomposed as
\begin{equation}
 \eta = 
\begin{pmatrix}
 1\\
 0
\end{pmatrix}
\otimes\theta(x) \otimes \ep(y), 
\end{equation}
where $\theta(x)$ ($\ep(y)$) is a three- (seven-)dimensional spinor satisfying
the Majorana conditions
\begin{eqnarray}
\theta^*&=&\theta,\\
\epsilon^*&=&C_7\epsilon.\label{eq:mcseven} 
\end{eqnarray}
The three-dimensional Killing spinor equation,
\begin{equation}
 \nabla^{(3)}_{\mu}\theta = \frac{a}{2}\tilde{\gamma}_{\mu}\theta,
\label{eq:adstheta}
\end{equation}
is also assumed, where $\nabla^{(3)}$ is the Levi-Civita connection of $AdS_{3}$ and $a$ is a constant
related to the three-dimensional cosmological constant as $\Lambda=-2a^2$.

\subsection{Decomposition of the Killing spinor equations}

By using this ansatz, the Killing spinor equations can be written as
\begin{eqnarray}
\nabla^{(-)}_{m}\epsilon&=&0,\label{eq:diffcond}\\
	\left(a -
	ie^{-A}(\del_{m}A)\gamma^{m} -
	\frac{1}{2}e^{-A}h\right)\epsilon&=&0,
	\label{eq:delpsi3}\\
	\left(-2\del_{m}\Phi\gamma^{m} +
	\frac{1}{3!}H_{mnp}\gamma^{mnp} -
	ih\right)\epsilon&=&0,
	\label{eq:dellambda}\\
	F_{mn}\gamma^{mn}\epsilon&=&0.
	\label{eq:delchi}
\end{eqnarray}
The seven-dimensional covariant derivative\footnote{
We omit the superscript $(7)$ for simplicity.}
$\nabla^{(-)}_{m}$ is defined by
\begin{eqnarray}
 \nabla^{(-)}_{m}\epsilon &\equiv& \nabla^{}_{m}\epsilon 
- \frac{1}{4\cdot 2!}H_{mnp}\gamma^{np}\epsilon,\\
&=&\del_{m}\epsilon +
\frac{1}{4}\omega_{mnp}\gamma^{np}\epsilon
- \frac{1}{4\cdot 2!}H_{mnp}\gamma^{np}\epsilon,
\end{eqnarray}
where $\omega_{mnp}=\omega_{m\hat{n}\hat{p}}e^{\hat{n}}_ne^{\hat{p}}_p$
is the spin connection and the three-form flux $H_{nmp}$ is interpreted as a torsion.
It is useful to define a nonminimal spin connection as 
$\omega^{(-)}_{mnp}=\omega_{mnp}-\frac{1}{2}H_{mnp}$, which allows that the covariant derivative
$\nabla^{(-)}_m$ can be written in the conventional form as
\begin{equation}
 \nabla^{(-)}_m\ep=\del_m\epsilon+\frac{1}{4}\omega^{(-)}_{mnp}\gamma^{np}\ep.
\end{equation}

\section{Killing spinors and their bilinear forms}
\label{sec:killing}

In this section, we first introduce the fermion representation of seven-dimensional 
spinors, which is useful for solving Killing spinor equations explicitly.
We simplify the forms of the Killing spinors, for the cases $N=1,2,3$, and $4$,
using the differential Killing spinor equation (\ref{eq:diffcond}) and 
the degrees of freedom of the local Lorentz transformation.
Accordingly, the differential Killing spinor equation can be solved by 
setting some components, irreducible representations of $G$-structure, 
of the (nonminimal) spin connection equal to zero.
Then, we consider the bilinear $n$-forms of the Killing spinors,
\begin{equation}
\frac{1}{n!}\left(\ep^\dag_{a}\gamma_{\hat{m}_{1}\hat{m}_{2}\dots \hat{m}_{n}}\ep_{b}
\right)e^{\hat{m}_1\hat{m}_2\cdots\hat{m}_n},
\end{equation}
where $n=1,2,3$\footnote{
The bilinear $n$-forms defined similarly for $n\ge4$ are not independent but Hodge dual
of those with $n\le3$.} and $a,b=1,\cdots,N$,
which characterize the $G$-structure
of the manifold $\mathcal{M}_7$ described by the solutions of the Killing spinor equations.

\subsection{Fermion representation of seven-dimensional spinors}
\label{sec:fermion}

A fermion representation of seven-dimensional ($SO(7)$) spinor
can be obtained using the following fermionic creation and annihilation operators
satisfying $\{a^i,a^{\bar{j}}\}=\delta^{i\bar{j}}$:
\begin{eqnarray}
a^{i} &\equiv& \frac{1}{2}(\gamma^{\hat{2i-1}} - i\gamma^{\hat{2i}}),\\
a^{\bar{i}} &=& \frac{1}{2}(\gamma^{\hat{2i-1}} + i\gamma^{\hat{2i}})=(a^i)^\dag,
\end{eqnarray}
where $i=1,2,3$. The seven-dimensional gamma matrices 
can be represented using these operators as
\begin{eqnarray}
 \gamma^{\hat{1}}&=&a^1+a^{\bar{1}},\qquad
\gamma^{\hat{2}}=i(a^1-a^{\bar{1}}),\label{eq:gamma1}\\
 \gamma^{\hat{3}}&=&a^2+a^{\bar{2}},\qquad
\gamma^{\hat{4}}=i(a^2-a^{\bar{2}}),\label{eq:gamma2}\\
 \gamma^{\hat{5}}&=&a^3+a^{\bar{3}},\qquad
\gamma^{\hat{6}}=i(a^3-a^{\bar{3}}),\label{eq:gamma3}\\
\gamma^{\hat{7}}&=&(1-2a^{\bar{1}}a^1)(1-2a^{\bar{2}}a^2)(1-2a^{\bar{3}}a^3).\label{eq:gamma4}
\end{eqnarray}
Here, we take $y^7$-direction as a special reference direction. The creation and annihilation
operators $(a^i,a^{\bar{i}})$ belong to $(3,\bar{3})$ representations of $SU(3)\subset SO(6)$,
where this $SO(6)$ is the rotation group orthogonal to the $y^7$-direction.

In this representation, the charge conjugation matrix $C_7$ (\ref{eq:ccseven}) becomes 
\begin{equation}
C_7=(a^1-a^{\bar{1}})(a^2-a^{\bar{2}})(a^3-a^{\bar{3}}). 
\end{equation}

We can represent an $SO(7)$ spinor as a linear combination of Fock states on
the (normalized) Clifford vacuum $\ket{0}$ defined by
\begin{equation}
 a^{i}\ket{0}=0, \quad \textrm{for} \quad i=1,2,3.
\end{equation}
The most general $SO(7)$ spinor, therefore, can be written as
\begin{equation}
 \ep = \ket{0}M + \ket{i}N_{\bar{i}}
	+\frac{1}{2} \ket{ij}P_{\bar{i}\bar{j}}
	+ \ket{123}Q,\label{eq:spinor}
\end{equation}
where we denote $\ket{ij\cdots k}\equiv a^{\bar{i}}a^{\bar{j}}\cdots a^{\bar{k}}\ket{0}$.
In this representation, the conjugate operations on the spinor are realized by
\begin{eqnarray}
 \ep^*&=& \ket{0}M^* + \ket{i}(N_{\bar{i}})^*
	+ \frac{1}{2}\ket{ij}(P_{\bar{i}\bar{j}})^*
	+ \ket{123}Q^*,\\
 \ep^\dag&=&M^*\bra{0} + (N_{\bar{i}})^*\bra{i}
	+ \frac{1}{2}(P_{\bar{i}\bar{j}})^*\bra{ij}
	+ Q^*\bra{123},
\end{eqnarray}
where $\bra{ij\cdots k}=\bra{0}a^k\cdots a^ja^i$.
The Majorana condition (\ref{eq:mcseven}), in this representation, is written as
\begin{eqnarray}
M^*&=&-Q,\label{eq:maj1}\\
(N_{\bar{i}})^*&=&\frac{1}{2}\epsilon_{ijk}P_{\bar{j}\bar{k}}\label{eq:maj2}.
\end{eqnarray}

In general, the form of a spinor can be simplified by choosing a special
local Lorentz frame, in which we can analyze the Killing spinor 
equations systematically \cite{Gillard:2004xq}.
In this paper, we consider the cases that the numbers of independent Killing spinors
are $N=1,2,3$, and $4$. These four independent Majorana spinors 
can be expressed using the basis spinors
\begin{eqnarray}
 \ep_{1} &\equiv& \frac{1}{\sqrt{2}}(\ket{0} - \ket{123}),\label{eq:epsilon1}\\
 \ep_{2} &\equiv& \frac{i}{\sqrt{2}}(\ket{0} + \ket{123}),\\
 \ep_{3} &\equiv& \frac{1}{\sqrt{2}}(\ket{3} + \ket{12}),\\
 \ep_{4} &\equiv& \frac{i}{\sqrt{2}}(\ket{3} - \ket{12}),
\end{eqnarray}
satisfying $\epsilon^\dag_i\epsilon_j=\delta_{ij}$.

\subsection{The differential Killing spinor equation}

Let us first investigate the differential Killing spinor equation $(\ref{eq:diffcond})$ 
for $N=1,2,3$, and $4$.

\subsubsection{$N=1$ case}

For $N=1$ case, we can take the Killing spinor as
\begin{equation}
 \epsilon=\alpha\epsilon_1,
\end{equation}
where $\alpha=\alpha(y)$ is a real function. From (\ref{eq:diffcond}),
however, 
\begin{eqnarray}
 0&=&(\nabla^{(-)}_m\epsilon)^\dag\epsilon
+\epsilon^\dag\nabla^{(-)}_m\epsilon,\\
&=&\partial_m(\epsilon^\dag\epsilon),\\
&=&\partial_m\alpha^2,
\end{eqnarray}
thus, the $\alpha$ must be a constant. Thus, we normalize 
$\epsilon$ as $\epsilon^\dag\epsilon=1$. Then the first Killing equation
(\ref{eq:diffcond}) 
restricts the spin connection to
\begin{equation}
(\omega^{(-)}_m)_{nl}=(\omega^{(-)}_m)^{(14)}_{nl}, \label{eq:diffsol1}
\end{equation}
where the decomposition of spin connection is defined in Appendix \ref{sec:g2}.
In other words, only the $(\omega_m^{(-)})^{(14)}_{nl}$ component of the spin connection 
is allowed to be nonvanishing.
We must also note that Eq. (\ref{eq:diffsol1})
holds only in this special local Lorentz frame, since the spin connection is transformed 
inhomogeneously under the local Lorentz transformation.

\subsubsection{$N=2$ case}

For $N=2$, two (Majorana) Killing spinors can be taken as $\ep_{1}$ and
\begin{equation}
 \ep = \alpha\ep_{1} + \beta\ep_{2},
\end{equation}
where $\alpha = \alpha(y)$ and $\beta = \beta(y)$ are real functions satisfying
$\alpha^2+\beta^2=$ constant, from the similar argument with the $N=1$ case. 
From (\ref{eq:diffcond}), in this case,
\begin{equation}
0=\nabla^{(-)}_m\epsilon=(\partial_m\alpha)\epsilon_1+(\partial_m\beta)\epsilon_2+
\beta\nabla^{(-)}_m\epsilon_2.\label{eq:twoks}
\end{equation}
By multiplying (\ref{eq:twoks}) by $\epsilon^\dag_1$,
we can obtain $\partial_m\alpha=0$, thus both $\alpha$ and $\beta$ must be constants, since
\begin{equation}
 0=\partial_m(\epsilon_1^\dag\epsilon_2)
=\epsilon_1^\dag\nabla^{(-)}_m\epsilon_2.
\end{equation}

Therefore, we can choose  the two Killing spinors as $\epsilon_1$ and $\epsilon_2$
without the loss of generality. The first Killing spinor equation (\ref{eq:diffcond}) then yields
\begin{eqnarray}
(\omega^{(-)}_m)_{nl}&=&(\omega^{(-)}_m)^{(8)}_{nl},\label{eq:diffsol2}
\end{eqnarray}
where the decomposition is defined in Appendix \ref{sec:su3}.
That is, only the $(\omega_m^{(-)})^{(8)}_{nl}$ component is allowed to be nonvanishing.
It must again be noted that Eq. (\ref{eq:diffsol2}) 
only hold in this special local Lorentz frame.

\subsubsection{$N = 3$ and $4$ cases}

Similarly, for $N=3$ and $4$ cases, the third and fourth 
Killing spinors can be 
taken to be $\epsilon_i$ $(i=1,2,3)$ and $\epsilon_i$ $(i=1,2,3,4)$ respectively.
For both cases,
the first Killing spinor equation (\ref{eq:diffcond}) gives the same conditions,
in this special local Lorentz frame, as
\begin{eqnarray}
(\omega^{(-)}_m)_{nl}&=&(\omega^{(-)}_m)^{(3)}_{nl},
\end{eqnarray}
where the decomposition of the spin connection is defined in Appendix \ref{sec:su2}.

\subsection{Bilinear forms of the Killing spinors}\label{sec:bilinear}

Using the Killing spinors, we can construct bilinear forms characterizing
the G-structure. At the end of this section, we consider them for $N=1,2,3$, 
and $4$ cases, respectively.

\subsubsection{$N=1$ case}

For $N = 1$, only one nontrivial spinor bilinear $\sigma$,
\begin{eqnarray}
 \sigma &\equiv& -\frac{i}{3!}\ep_{1}^{\dagger}\gamma_{\hat{m}\hat{n}\hat{p}}\ep_{1}e^{\hat{m}\hat{n}\hat{p}},\\
 &=& - e^{\hat{2}\hat{4}\hat{6}} + 
e^{\hat{2}\hat{3}\hat{5}} + 
e^{\hat{1}\hat{4}\hat{5}} + 
e^{\hat{1}\hat{3}\hat{6}} - 
e^{\hat{1}\hat{2}\hat{7}} - 
e^{\hat{3}\hat{4}\hat{7}} - 
e^{\hat{5}\hat{6}\hat{7}},
\end{eqnarray}
can be obtained as the bilinear of the Killing spinor,
where $e^{\hat{m}\hat{n}\hat{p}}=e^{\hat{m}}\wedge e^{\hat{n}}\wedge e^{\hat{p}}$ 
denotes the wedge product of vierbein one-forms of $\mathcal{M}_{7}$.
This three-form and its Hodge dual $(\sigma,*\sigma)$ define a $G_{2}$-structure on 
$\mathcal{M}_{7}$ \cite{Chiossi:2002tw, Bryant:2005mz}.

\subsubsection{$N = 2$ case}

For $N=2$ case, one can obtain a one-form, 
\begin{eqnarray}
 K &\equiv& -\frac{i}{2}\left(\ep_{1}^{\dagger}\gamma_{\hat{m}}\ep_{2}
-\ep_{2}^{\dagger}\gamma_{\hat{m}}\ep_{1}\right)e^{\hat{m}} \nonumber\\
&=& e^{\hat{7}},
\end{eqnarray}
a two-form, 
\begin{eqnarray}
 J &\equiv& \frac{1}{4}\left(\ep_{1}^{\dagger}\gamma_{\hat{m}\hat{n}}\ep_{2}
-\ep_{2}^{\dagger}\gamma_{mn}\ep_{1}\right)e^{\hat{m}\hat{n}} \nonumber\\
  &=& e^{\hat{1}\hat{2}}+e^{\hat{3}\hat{4}}+e^{\hat{5}\hat{6}},
\end{eqnarray}
and three three-forms,
\begin{eqnarray}
\sigma_1&=&-\frac{i}{6}\epsilon_1^\dag\gamma_{\hat{m}\hat{n}\hat{p}}\epsilon_1e^{mnp},\nonumber\\
&=&-e^{\hat{1}\hat{2}\hat{7}}-e^{\hat{3}\hat{4}\hat{7}}
-e^{\hat{5}\hat{6}\hat{7}}+e^{\hat{2}\hat{3}\hat{5}}
+e^{\hat{1}\hat{4}\hat{5}}+e^{\hat{1}\hat{3}\hat{6}}
-e^{\hat{2}\hat{4}\hat{6}},\label{eq:n23form1}\\
\sigma_2&=&-\frac{i}{6}\epsilon_2^\dag\gamma_{\hat{m}\hat{n}\hat{p}}\epsilon_2e^{mnp},\nonumber\\
&=&-e^{\hat{1}\hat{2}\hat{7}}-e^{\hat{3}\hat{4}\hat{7}}
-e^{\hat{5}\hat{6}\hat{7}}-e^{\hat{2}\hat{3}\hat{5}}
-e^{\hat{1}\hat{4}\hat{5}}-e^{\hat{1}\hat{3}\hat{6}}+e^{\hat{2}\hat{4}\hat{6}},\label{eq:n23form2}\\
\sigma_3&=&-\frac{i}{12}\left(
\epsilon_1^\dag\gamma_{\hat{m}\hat{n}\hat{p}}\epsilon_2+
\epsilon_2^\dag\gamma_{\hat{m}\hat{n}\hat{p}}\epsilon_1\right)e^{mnp},\nonumber\\
&=&-e^{\hat{1}\hat{3}\hat{5}}+e^{\hat{1}\hat{4}\hat{6}}
+e^{\hat{2}\hat{3}\hat{6}}+e^{\hat{2}\hat{4}\hat{5}},\label{eq:n23form3}
\end{eqnarray}
as the bilinear of the two Killing spinors.

Using the differential Killing spinor equation (\ref{eq:diffcond}),
one can show that $K=K_m dy^m$ satisfies the Killing equation
\begin{equation}
 \nabla_m K_n+\nabla_n K_m=0.
\end{equation}
Then, by choosing the $y^7$ coordinate along the direction of the Killing vector, 
the metric can be written in the form
\begin{equation}
 \d s_{\mathcal{M}_{7}}^{2} = g_{\tilde{m}\tilde{n}}\d y^{\tilde{m}} \d y^{\tilde{n}} + (dy^7+w)^2,
\label{eq:n2metric}
\end{equation}
where $g_{\tilde{m}\tilde{n}}$ and 
$w=w_{\tilde{m}}dy^{\tilde{m}}\ (\tilde{m},\tilde{n} =1,\cdots,6)$,
both of which are independent of $y^7$, are the metric of the six-dimensional
submanifold $\mathcal{M}_6$ and a one-form on it.
The three-forms (\ref{eq:n23form1})--(\ref{eq:n23form3}) can be written 
using $K$, $J$, and the three-form
\begin{equation}
 \Omega=(e^{\hat{1}}+ie^{\hat{2}})\wedge(e^{\hat{3}}+ie^{\hat{4}})\wedge(e^{\hat{5}}+ie^{\hat{6}}),
\end{equation}
as
\begin{eqnarray}
\sigma_1&=&-J\wedge K+\textrm{Im}\ \! \Omega,\\
\sigma_2&=&-J\wedge K-\textrm{Im}\ \! \Omega,\\
\sigma_3&=&-\textrm{Re}\ \!\Omega.
\end{eqnarray}
The independent forms $(K,J,\Omega)$ define the $SU(3)$-structure
on $\mathcal{M}_7$ \cite{Chiossi:2002tw, Cardoso:2002hd}.

\subsubsection{$N = 3$ and $4$ cases}

For $N=3$ case, nonvanishing bilinears of three Killing spinors are
three one-forms, 
\begin{eqnarray}
K_1&=&-\frac{i}{2}\left(\epsilon^\dag_1\gamma_{\hat{m}}\epsilon_2
-\epsilon_2^\dag\gamma_{\hat{m}}\epsilon_1\right)e^{\hat{m}}=e^{\hat{7}},\label{eq:n31form1}\\
K_2&=&\frac{i}{2}\left(\epsilon^\dag_2\gamma_{\hat{m}}\epsilon_3
-\epsilon_3^\dag\gamma_{\hat{m}}\epsilon_2\right)e^{\hat{m}}=e^{\hat{5}},\label{eq:n31form2}\\
K_3&=&\frac{i}{2}\left(\epsilon^\dag_3\gamma_{\hat{m}}\epsilon_1
-\epsilon_1^\dag\gamma_{\hat{m}}\epsilon_3\right)e^{\hat{m}}=e^{\hat{6}},\label{eq:n31form3}
\end{eqnarray}
three two-forms,
\begin{eqnarray}
J_1&=&\frac{1}{4}\left(
\epsilon^\dag_1\gamma_{\hat{m}\hat{n}}\epsilon_2
-\epsilon^\dag_2\gamma_{\hat{m}\hat{n}}\epsilon_1\right)e^{\hat{m}\hat{n}},\label{eq:n32form1}\\
J_2&=&-\frac{1}{4}\left(
\epsilon^\dag_2\gamma_{\hat{m}\hat{n}}\epsilon_3
-\epsilon^\dag_3\gamma_{\hat{m}\hat{n}}\epsilon_2\right)e^{\hat{m}\hat{n}},\label{eq:n32form2}\\
J_3&=&\frac{1}{4}\left(
\epsilon^\dag_3\gamma_{\hat{m}\hat{n}}\epsilon_1
-\epsilon^\dag_3\gamma_{\hat{m}\hat{n}}\epsilon_3\right)e^{\hat{m}\hat{n}}, \label{eq:n32form3}
\end{eqnarray}
and six three-forms,
\begin{eqnarray}
 \sigma_1&=&-\frac{i}{3!}(\epsilon^\dag_1\gamma_{\hat{m}\hat{n}\hat{p}}\epsilon_1)
e^{\hat{m}\hat{n}\hat{p}},\label{eq:n33form1}\\
 \sigma_2&=&-\frac{i}{3!}(\epsilon^\dag_2\gamma_{\hat{m}\hat{n}\hat{p}}\epsilon_2)
e^{\hat{m}\hat{n}\hat{p}},\label{eq:n33form2}\\
 \sigma_3&=&-\frac{i}{3!}\frac{1}{2}\left(
\epsilon^\dag_1\gamma_{\hat{m}\hat{n}\hat{p}}\epsilon_2
+\epsilon^\dag_2\gamma_{\hat{m}\hat{n}\hat{p}}\epsilon_1
\right)e^{\hat{m}\hat{n}\hat{p}},\label{eq:n33form3}\\
 \sigma_4&=&-\frac{i}{3!}(\epsilon^\dag_3\gamma_{\hat{m}\hat{n}\hat{p}}\epsilon_3)
e^{\hat{m}\hat{n}\hat{p}},\label{eq:n33form4}\\
 \sigma_5&=&-\frac{i}{3!}\frac{1}{2}\left(
\epsilon^\dag_2\gamma_{\hat{m}\hat{n}\hat{p}}\epsilon_3
+\epsilon^\dag_3\gamma_{\hat{m}\hat{n}\hat{p}}\epsilon_2
\right)e^{\hat{m}\hat{n}\hat{p}},\label{eq:n33form5}\\
 \sigma_6&=&-\frac{i}{3!}\frac{1}{2}\left(
\epsilon^\dag_3\gamma_{\hat{m}\hat{n}\hat{p}}\epsilon_1
+\epsilon^\dag_1\gamma_{\hat{m}\hat{n}\hat{p}}\epsilon_3
\right)e^{\hat{m}\hat{n}\hat{p}}.\label{eq:n33form6}
\end{eqnarray}

Similarly to the $N=2$ case, one can obtain three Killing vectors from the one-forms
(\ref{eq:n31form1})--(\ref{eq:n31form3}):
\begin{equation}
  \nabla_m (K_I)_n+\nabla_n (K_I)_m=0. \qquad (I=1,2,3)
\end{equation}
By choosing the $(y^5,y^6,y^7)$ coordinates along the directions of these three 
Killing vectors $(K_2,K_3,K_1)$, the metric can be written as\footnote{See also Ref.~\citen{Fernandez:2008aa}.}
\begin{equation}
 \d s_{\mathcal{M}_{7}}^{2}
	=  g_{mn}\d y^{m} \d y^{n}
	+ (\d y^{5}+w_{2})^2
	+ (\d y^{6}+w_{3})^2
	+ (\d y^{7}+w_{1})^2,\label{eq:n3metric}
\end{equation}
where $g_{\tilde{m}\tilde{n}}$, $(\tilde{m},\tilde{n}=1,\cdots,4)$
and $w_{I}=(w_I)_{\tilde{m}}dy^{\tilde{m}}$,
all of which are independent of $y^5$, $y^6$, and $y^7$, 
are the metric of four-dimensional submanifold $\mathcal{M}_{4}$ and one-forms on it.
The two-forms (\ref{eq:n32form1})--(\ref{eq:n32form3}) and the three-forms 
(\ref{eq:n33form1})--(\ref{eq:n33form6}) can be rewritten as
\begin{eqnarray}
 J_1&=&J+K_2\wedge K_3,\\
 J_2&=&\textrm{Im}\ \!\Omega+K_3\wedge K_1,\\
 J_3&=&\textrm{Re}\ \!\Omega-K_1\wedge K_2,\\
 \sigma_1&=&-J\wedge K_1+\Im\Omega\wedge K_2
+\Re\Omega\wedge K_2-K_1\wedge K_2\wedge K_3,\\
 \sigma_2&=&-J\wedge K_1-\Im\Omega\wedge K_2
-\Re\Omega\wedge K_2-K_1\wedge K_2\wedge K_3,\\
 \sigma_3&=&-\Re\Omega\wedge K_2+\Im\Omega\wedge K_3,\\
 \sigma_4&=&J\wedge K_1-\Im\Omega\wedge K_2
+\Re\Omega\wedge K_2-K_1\wedge K_2\wedge K_3,\\
 \sigma_5&=&-J\wedge K_3+\Re\Omega\wedge K_1,\\
 \sigma_6&=&-\Im\Omega\wedge K_1-J\wedge K_2,
\end{eqnarray}
by means of
\begin{eqnarray}
J&=&e^{\hat{1}\hat{2}}+e^{\hat{3}\hat{4}},\\
\Omega&=&(e^{\hat{1}}+ie^{\hat{2}})\wedge(e^{\hat{3}}+ie^{\hat{4}}).
\end{eqnarray}
The independent forms $(K_I,J,\Omega)$ 
define an $SU(2)$-structure on $\mathcal{M}_{7}$.

For $N=4$ case, additional three one-forms
\begin{eqnarray}
K_4&=&\frac{i}{2}\left(\epsilon^\dag_3\gamma_{\hat{m}}\epsilon_4
-\epsilon_4^\dag\gamma_{\hat{m}}\epsilon_3\right)e^{\hat{m}}=e^{\hat{7}}=K_1,\\
K_5&=&\frac{i}{2}\left(\epsilon^\dag_4\gamma_{\hat{m}}\epsilon_1
-\epsilon_1^\dag\gamma_{\hat{m}}\epsilon_4\right)e^{\hat{m}}=e^{\hat{5}}=K_2,\\
K_6&=&-\frac{i}{2}\left(\epsilon^\dag_2\gamma_{\hat{m}}\epsilon_4
-\epsilon_4^\dag\gamma_{\hat{m}}\epsilon_2\right)e^{\hat{m}}=e^{\hat{6}}=K_3,
\end{eqnarray}
three two-forms
\begin{eqnarray}
J_4&=&\frac{1}{4}\left(
\epsilon^\dag_3\gamma_{\hat{m}\hat{n}}\epsilon_4
-\epsilon^\dag_4\gamma_{\hat{m}\hat{n}}\epsilon_3\right)e^{\hat{m}\hat{n}},\\
&=&J-K_2\wedge K_3,\\
J_5&=&-\frac{1}{4}\left(
\epsilon^\dag_4\gamma_{\hat{m}\hat{n}}\epsilon_1
-\epsilon^\dag_1\gamma_{\hat{m}\hat{n}}\epsilon_4\right)e^{\hat{m}\hat{n}},\\
&=&-\textrm{Im}\ \!\Omega+K_3\wedge K_1,\\
J_6&=&\frac{1}{4}\left(
\epsilon^\dag_2\gamma_{\hat{m}\hat{n}}\epsilon_4
-\epsilon^\dag_4\gamma_{\hat{m}\hat{n}}\epsilon_2\right)e^{\hat{m}\hat{n}}, \\
&=&\textrm{Re}\ \!\Omega+K_1\wedge K_2,
\end{eqnarray}
and four three-forms
\begin{eqnarray}
\sigma_7&=&-\frac{i}{3!}(\epsilon^\dag_4\gamma_{\hat{m}\hat{n}\hat{p}}\epsilon_4)
e^{\hat{m}\hat{n}\hat{p}},\\
&=&J\wedge K_1+\Im\Omega\wedge K_2-\Re\Omega\wedge K_3-K_1\wedge K_2\wedge K_3,\\
\sigma_8&=&-\frac{i}{3!}\frac{1}{2}\left(
\epsilon^\dag_3\gamma_{\hat{m}\hat{n}\hat{p}}\epsilon_4
+\epsilon^\dag_4\gamma_{\hat{m}\hat{n}\hat{p}}\epsilon_3
\right)e^{\hat{m}\hat{n}\hat{p}},\\
&=&\Re\Omega\wedge K_2+\Im\Omega\wedge K_3,\\
\sigma_9&=&-\frac{i}{3!}\frac{1}{2}\left(
\epsilon^\dag_4\gamma_{\hat{m}\hat{n}\hat{p}}\epsilon_1
+\epsilon^\dag_1\gamma_{\hat{m}\hat{n}\hat{p}}\epsilon_4
\right)e^{\hat{m}\hat{n}\hat{p}},\\
&=&-J\wedge K_3+\Re\Omega\wedge K_1,\\
\sigma_{10} &=&-\frac{i}{3!}\frac{1}{2}\left(
\epsilon^\dag_2\gamma_{\hat{m}\hat{n}\hat{p}}\epsilon_4
+\epsilon^\dag_4\gamma_{\hat{m}\hat{n}\hat{p}}\epsilon_2
\right)e^{\hat{m}\hat{n}\hat{p}},\\
&=&\Im\Omega\wedge K_1-J\wedge K_2,\\
\end{eqnarray}
constructed from the fourth Killing spinor $\epsilon_4$,
do not yield any new independent forms.
Thus $\mathcal{M}_{7}$ is again characterized by the $SU(2)$-structure 
given by $(K_I,J,\Omega)$.

\section{Solutions of Killing spinor equations}
\label{sec:results}

In this section, we solve the remaining algebraic Killing spinor equations 
(\ref{eq:delpsi3})--(\ref{eq:delchi}) for the cases in which there are $N=1$, $2$, $3$, 
and $4$ Killing spinors. We must be careful, however, that 
the conditions on the spin connections hold only in this special local Lorentz frame.

\subsection{Solutions of algebraic equations}

Let us begin with solving the first algebraic equation (\ref{eq:delpsi3}).
We show that the warp factor $A$ must be constant, independent of the number $N$ of Killing spinors.

For $N=1$, Eq. (\ref{eq:delpsi3}) must hold
on $\epsilon_1$;
\begin{equation}
 	\left(\left(2a -e^{-A}h\right)
	-2ie^{-A}(\del_{m}A)\gamma^{m} 
	\right)\epsilon_1=0.\label{eq:killing2}
\end{equation}
By multiplying $\ep_{1}^{\dagger}$ or
$\ep_1^\dag\gamma_n$, we obtain
\begin{eqnarray}
 2a-e^{-A}h&=&0,\label{eq:delpsi3-5}\\
 \partial_mA&=&0.
\end{eqnarray}
The warp factor $A$, therefore, must be constant.  We can set $A=0$
by rescaling the three dimensional metric, which is equivalent to rescaling
$a$ into $ae^{-A}$. 
We finally obtain
\begin{eqnarray}
 h&=&2a,\\
 A &=& 0.
\end{eqnarray}
Then the Killing spinor equation (\ref{eq:delpsi3}) identically holds.
No additionalcondition is required for $N\ge2$ cases. 

The remaining algebraic equations (\ref{eq:dellambda}) and (\ref{eq:delchi}) 
can be solved using the fermion representation introduced above. 
Here, we explicitly show how to solve them by using an example for 
(\ref{eq:delchi}).\footnote{See Refs.~\citen{Gillard:2004xq} and \citen{Gran:2006pe} for details of this technique.}
Then the solutions for all the Killing spinor equations are summarized for 
the cases of $N=1, 2, 3$ and $4$, respectively.

For $\ep_{1}$ (\ref{eq:epsilon1}), Eq. (\ref{eq:dellambda}) is rewritten as
\begin{equation}
 F_{mn}\gamma^{mn}\ep_{1} = 
\frac{1}{\sqrt{2}}F_{mn}\gamma^{mn}(\ket{0} - \ket{123})
= 0.
\end{equation}
Using fermion representation introduced in \S \ref{sec:fermion},
we can compute
\begin{eqnarray}
\frac{1}{2}F_{mn}\gamma^{mn}\ket{0} 
  &=& F_{i\bar{j}}\delta^{i\bar{j}}\ket{0}
  + F_{\bar{i}\bar{j}}\ket{ij}
  + \sqrt{2}F_{\bar{i}7}\ket{i}, \\
 \frac{1}{2}F_{mn}\gamma^{mn}\ket{123}
  &=& -\ep_{\bar{i}\bar{j}\bar{k}}F_{jk}\ket{i}
  - F_{i\bar{j}}\delta^{i\bar{j}}\ket{123}
  - \frac{1}{\sqrt{2}}\ep_{\bar{i}\bar{j}\bar{k}}F_{k7}\ket{ij},
\end{eqnarray}
where components $F_{i\bar{j}}$ and $F_{ij}$ are defined in Appendix \ref{sec:decomp}. 
We obtain
\begin{eqnarray}
&&F_{i\bar{j}}\delta^{i\bar{j}}\ket{0}
 + \sqrt{2}(F_{\bar{i}7} + \frac{1}{\sqrt{2}} \ep_{\bar{i}\bar{j}\bar{k}}F_{jk})\ket{i}\nn
&&\hspace{16mm}
 + (F_{\bar{i}\bar{j}} +\frac{1}{\sqrt{2}}\ep_{\bar{i}\bar{j}\bar{k}}F_{k7})\ket{ij}
 + F_{i\bar{j}}\delta^{i\bar{j}}\ket{123}=0.
\end{eqnarray}
Since each Fock state is independent, this is equivalent to
\begin{eqnarray}
 F_{i\bar{j}}\delta^{i\bar{j}} &=& 0, \\
  F_{\bar{i}7} + \frac{1}{\sqrt{2}}\ep_{\bar{i}\bar{j}\bar{k}}F_{jk}&=& 0.
\end{eqnarray}

The other equations can be solved in a similar manner, 
using irreducible decompositions by G-structure given 
in Appendix \ref{sec:decomp}. 

\subsection{Solutions of the Killing spinor equations} 

We summarize the solutions of all the Killing spinor 
equations as follows in the cases of $N=1,2,3$ and $4$.

\subsubsection{For $N = 1$}

For the $N=1$ case, $\mathcal{M}_7$ has a $G_{2}$-structure, defined by $(\sigma,*\sigma)$, 
which decomposes
all the fluxes into its irreducible representations as explained in Appendix \ref{sec:g2}.
The solution of all the Killing spinor equations can be simply written as
\begin{eqnarray}
 A&=& 0, \\
d\Phi &=& -2H^{(7)},\\
 H &=& \frac{2}{7}a\sigma +H^{(7)}\lrcorner\ast\sigma + H^{(27)},\\
 F &=& F^{(14)},\\
 (\omm_{m})_{nl} &=& (\omm_{m})^{(14)}_{nl}. \label{eq:sol14}
\end{eqnarray}

The components $H^{(7)}$, $H^{(27)}$, $F^{(14)}$, $(\omega^{(-)}_{m})^{(14)}_{nl}$, and $a$ are not restricted.
Again, we would like to stress that the spin connection $\omega^{(-)}$ is not tensor
and the condition (\ref{eq:sol14}) holds only in this frame.

\subsubsection{For $N = 2$}

For the $N=2$ case, $\mathcal{M}_7$ has an $SU(3)$-structure defined by $(K,J,\Omega)$.
The solutions of all the Killing spinor equations are given by
\begin{eqnarray}
 A &=& 0,\\
 d\Phi &=& -i(H^{(3)} - H^{(\bar{3})}),\\
 H &=& -(d\Phi\lrcorner J)\wedge J + H^{(6)} + H^{(\bar{6})}
	+ \left(-\frac{2}{3}aJ + \tilde{H}^{(8)}\right)\wedge K,\\
 F &=& F^{(8)},\\
 (\omega^{(-)}_{m})_{nl} &=& (\omega^{(-)}_{m})^{(8)}_{nl}.
\end{eqnarray}

The components $H^{(3)}$, $H^{(6)}$, $\tilde{H}^{(8)}$, $F^{(8)}$,
$(\omega^{(-)}_{m})^{(8)}_{nl}$, and $a$ are not restricted.

\subsubsection{For $N = 3$}

For the $N=3$ case, we have an $SU(2)$-structure on $\mathcal{M}_7$ defined by $(K_I,J,\Omega)$.
The solutions of all the Killing spinor equations are
\begin{eqnarray}
 A &=& 0,\\
 d\Phi &=& \frac{i}{2}(H^{(2)}_{23} - H^{(\bar{2})}_{23}) 
	+ \frac{i}{2}(H^{(2)} - H^{(\bar{2})}),\\
 H &=& (H^{(2)} + H^{(\bar{2})})\wedge J \nn
&&	+ (H^{(1^{\prime})}_{1}J + H^{(3)}_{1})\wedge K^{1}
	+(H^{(1)}_{L}\bar{\Omega} + H^{(\bar{1})}_{L}\Omega + H^{(3)}_{L}) \wedge K^{L} \nn
&&	+ \frac{1}{2}(H^{(2)}_{IJ} + H^{(\bar{2})}_{IJ})\wedge K^{I}\wedge K^{J} \nn
&&	-2(a + H^{(1^{\prime})}_{1})K^{1}\wedge K^{2} \wedge K^{3},\\
 F &=& F^{(3)}, \\
 (\omega^{(-)}_{m})_{nl} &=& (\omega^{(-)}_{m})^{(3)}_{nl},
\end{eqnarray}
with additional conditions
\begin{eqnarray}
 H^{(2)}_{12} + iH^{(2)}_{31} &=& 0, \\
 H^{(2)}_{23} + \frac{i}{4}(H^{(\bar{2})}_{12} + iH^{(\bar{2})}_{31})\lrcorner\Omega &=& 0,\\
 H^{(1^{\prime})}_{1} + i(H^{(1)}_{2} - H^{(\bar{1})}_{2}) &=& 0, \\
 H^{(1)}_{2} + iH^{(1)}_{3} &=& 0,
\end{eqnarray}
where $I,J \in \{1,2,3\}$ and $L \in \{2,3\}$.

\subsubsection{For $N = 4$}

For the $N=4$ case, there is no additional independent bilinear form to the $N=3$ case.
The geometry of $\mathcal{M}_7$ admits again the $SU(2)$-structure.
However, the solutions of all the Killing spinor equations are more restrictive as
\begin{eqnarray}
 A &=& 0, \\
 d\Phi &=& \frac{i}{2}(H^{(2)} - H^{(\bar{2})}), \\
 H &=& -2d\Phi\lrcorner J \wedge J + H^{(3)}_{I}\wedge K^{I}
-2a K^{1}\wedge K^{2} \wedge K^{3}, \\
 F &=& F^{(3)}, \\
 (\omega^{(-)}_{m})_{nl} &=& (\omega^{(-)}_{m})^{(3)}_{nl} 
\end{eqnarray}
with no restrictions to $H^{(2)}$, $H^{(3)}_{I}$, $F^{(3)}$, $(\omega^{(-)}_{m})^{(3)}_{nl}$ and $a$.

It should be noted that there is no extra condition on the gauge field $F$ in addition
to the $N=3$ case. 

\section{$G$-structure and its torsion}
\label{sec:g-structure}

When there are no fluxes, $H=0$ and $F=0$, all the bilinear forms obtained in \S \ref{sec:bilinear} 
are closed. Then they define a familiar geometric structure on the special holonomy manifolds.
For example, the closed two-form $J$ with $dJ=0$ defines the complex
structure and becomes a K\"ahler form.\footnote{
Here we assume a hermitian metric.} If there are nontrivial fluxes,
however, the bilinear forms are no longer closed, but the deviation
from the closed forms is characterized as the torsion of the $G$-structure.
In this section, we compute these torsions from
the fluxes $H$ and $F$ using the Killing spinor equations. 
The $G$-structure is further classified in the class of torsion.

\subsection{Torsion class of the $G_2$-structure for $N = 1$}
\label{sec:g-1}

In the $N=1$ case, there is a $G_{2}$-structure \cite{Chiossi:2002tw, Bryant:2005mz} 
defined by $(\sigma,\ast\sigma)$. For general $G_2$-structure manifolds, these forms
$(\sigma,\ast\sigma)$ are not closed but classified by four torsion classes
$(\tau_0,\tau_1,\tau_2,\tau_3)$ as
\begin{eqnarray}
\d\sigma &=& \tau_{0}\ast\sigma + 3\tau_{1}\wedge\sigma + \ast \tau_{3},\\
\d\ast\sigma &=& 4\tau_{1}\wedge\ast\sigma + \tau_{2}\wedge\sigma.
\end{eqnarray}

On the other hand, using the Killing spinor equations, we can show that the bilinear forms 
$(\sigma,\ast\sigma)$ satisfy
\begin{eqnarray}
 \d\sigma &=& -6H^{(1)}\ast\sigma + 3H^{(7)}\wedge\sigma + \ast H^{(27)},\\
 \d\ast\sigma &=& 4H^{(7)}\wedge\ast\sigma.
\end{eqnarray}
Thus, the $G_2$-structure manifolds obtained by supersymmetric solutions
have the nontrivial torsion classes due to the three-form flux $H$ as\footnote{
Similar considerations were also given in Refs.~\citen{Friedrich:2001yp} and \citen{Friedrich:2001nh}.}
\begin{eqnarray}
 \tau_{0} &=& -6H^{(1)} = -\frac{6}{7}h, \\
 \tau_{1} &=& H^{(7)} = -\frac{1}{2}\d\Phi,\\
 \tau_{2} &=& 0,\\
 \tau_{3} &=& H^{(27)}.
\end{eqnarray}

\subsection{Torsion class of $SU(3)$-structure for $N = 2$}
\label{sec:g-2}

In the $N=2$ case, $\mathcal{M}_{7}$ admits $SU(3)$-structure \cite{Chiossi:2002tw}
characterized by the forms $(K,J,\Omega)$. For general $SU(3)$-structure manifolds,
these forms $(K,J,\Omega)$ are not closed but characterized by thirteen torsion classes as
\begin{eqnarray}
 \d K &=& V_{1}J + V_{2}\lrcorner\bar{\Omega} + \bar{V}_{2}\lrcorner\Omega + V_{3} + V_{4}\wedge K, \\
 \d J &=& \frac{3}{4}i
  \left(
   \mathcal{W}_{1}\bar{\Omega}
   - \bar{\mathcal{W}}_{1}\Omega
  \right)
  + \mathcal{W}_{3}
  + J\wedge \mathcal{W}_{4} \nn
  &&\hspace{1cm}
 + K \wedge
  \left[
   \frac{1}{3}(V_{5} + \bar{V_{5}})J
   + V_{6}\lrcorner\bar{\Omega}
   + \bar{V}_{6}\lrcorner\Omega
   + V_{7}
  \right], \\
 \d \Omega &=& \mathcal{W}_{1}J\wedge J + J\wedge\mathcal{W}_{2}
  + \Omega\wedge\mathcal{W}_{5} \nn
  && \hspace{3cm}
+ K\wedge
  \left[
   V_{5}\Omega - 4J\wedge \bar{V}_{6} + V_{8}
  \right].
\end{eqnarray}

We can show that the bilinear forms $(K,J,\Omega)$ of the Killing spinors
satisfy
\begin{eqnarray}
 \d K &=& \tilde{H}^{(1)}J+
\tilde{H}^{(3)}\lrcorner\bar{\Omega}+\tilde{H}^{(\bar{3})}\lrcorner\Omega+ \tilde{H}^{(8)},\\
 \d J &=& -3i\left(H^{(1)}\bar{\Omega}-H^{(\bar{1})}\Omega\right)-i\left(H^{(6)}-H^{(\bar{6})}\right)
\nn
&&\hspace{2cm}
-J\wedge i\left(H^{(3)}-H^{(\bar{3})}\right)
-2iK\wedge\left(\tilde{H}^{(\bar{3})}\lrcorner\bar{\Omega}
-\tilde{H}^{(3)}\lrcorner\Omega\right),\\
 \d \Omega &=& 4H^{(1)}J\wedge J-2i\Omega\wedge H^{(\bar{3})}
+K\wedge \left(3i\tilde{H}^{(1)}\Omega-8iJ\wedge\tilde{H}^{(3)}\right).
\end{eqnarray}
By substituting the solution of Killing spinor equations,
the torsion class of the $SU(3)$-structure is given by
\begin{eqnarray}
 V_{1} &=& \tilde{H}^{(1)}=-\frac{1}{3}h,\\
 V_{2} &=& \tilde{H}^{(3)}=0,\\
 V_{3} &=& \tilde{H}^{(8)},\\
 V_{4} &=& 0,\\
 V_{5} &=& 3i\tilde{H}^{(1)}=-ih,\\
 V_{6} &=& 2i\tilde{H}^{(3)}=0,\\
 V_{7} &=& 0,\\
 V_{8} &=& 0,
\end{eqnarray}
and
\begin{eqnarray}
 \mathcal{W}_{1} &=& -4H^{(1)}=0,\\
 \mathcal{W}_{2} &=& 0,\\
 \mathcal{W}_{3} &=& -i\left(H^{(6)}-H^{(\bar{6})}\right),\\
 \mathcal{W}_{4} &=& -i\left(H^{(3)} -H^{(\bar{3})}\right)=i\left(J\lrcorner d\Phi\right),\\
 \mathcal{W}_{5} &=& -2iH^{(\bar{3})}=2\left(\mathcal{P}^{(+)}\lrcorner d\Phi\right).
\end{eqnarray}

\subsection{Torsion class of $SU(2)$-structure for $N = 3$}
\label{sec:g-3}

In $N=3$ case, $\mathcal{M}_{7}$ admits $SU(2)$-structure,
defined by the forms $(K^I,J,\Omega)$. For general $SU(2)$-structure manifolds,
these forms $(K,J,\Omega)$ are characterized by seventy-five torsion
classes.
They consist of thirty singlets, thirty doublets, and fifteen triplets of $SU(2)$,
whose explicit forms are not given here. On the other hand, 
the exterior derivative of bilinear forms $(K,J,\Omega)$ can be computed 
using the Killing spinor equations as
\begin{eqnarray}
 \d K^{I} &=& H_{I}^{(1^{\prime})}J+ H_{I}^{(1)}\Omega + H_{I}^{(\bar{1})}\bar{\Omega} 
  + H_{I}^{(3)} \nn
&&  - \left(H_{IJ}^{(2)}+H_{IJ}^{(\bar{2})}\right)\wedge K^{J}  
+\frac{1}{2}\epsilon_{IJK}\tilde{H}^{(1)}K^{J}\wedge K^{K},\\
 \d J &=& i \left(H^{(2)}-H^{(\bar{2})} \right)\wedge J\nn
 &&
-2i
  \left(
   H_{I}^{(\bar{1})}\Omega - H_{I}^{(1)}\bar{\Omega}
  \right)\wedge K^{I} - \frac{i}{2}\left(
   H_{IJ}^{(2)} - H_{IJ}^{(\bar{2})}
  \right)\wedge K^{I} \wedge K^{J},\\
 \d\Omega
	&=&i H^{(\bar{2})}\wedge \Omega\nn
	&& +\left( 4iH_{I}^{(1)}J +2iH_{I}^{(1^{\prime})}\Omega\right)\wedge K^{I}
	+ \frac{1}{2}\left(H_{IJ}^{(\bar{2})}\lrcorner\Omega\right)\wedge K^{I}\wedge K^{J},
\end{eqnarray}
which reduce further by substituting the solutions.

\subsection{Torsion class of $SU(2)$-structure for $N = 4$}
\label{sec:g-4}

In $N=4$ case, $\mathcal{M}_{7}$ admits the same $SU(2)$-structure as in the $N=3$ case.
The Killing spinor equations, however, impose further constraints on the flux $H$.
Thus, the intrinsic torsion becomes simpler as
\begin{eqnarray}
 \d K^{I} &=& H_{I}^{(3)}
+\frac{1}{2}\epsilon_{IJK}\tilde{H}^{(1)}K^{J}\wedge K^{K},\\
 \d J &=& i \left(H^{(2)}-H^{(\bar{2})}  \right)\wedge J,\\
 \d\Omega &=& i H^{(\bar{2})}\wedge \Omega.
\end{eqnarray}

Thus far, we concentrate on analyzing the Killing spinor equations but do not 
take into account the equations of motion and the Bianchi identities.
In the next section, we show that it is sufficient to impose the Bianchi identities,
which imply the equations of motion automatically.

\section{Integrability conditions and equations of motion}
\label{sec:integrability}

Here, we show that the Killing spinor equations and Bianchi identities 
imply that all the equations of motion, including the leading $\alpha'$ correction,
are automatically satisfied.\footnote{
More precisely, the discussion in this section holds in the leading $\alpha'$ approximation
neglecting the $\alpha'^2$ order terms.}
This is well known in the case of $a=0$ and $R=0$  \cite{Gauntlett:2002sc}.
In the case of $R \neq 0$, however, the solutions of Killing spinor equations
and Bianchi identities is not always satisfy the Einstein equations
\cite{Fernandez:2008wa, Fernandez:2008aa, Fernandez:2008pf}.
In this paper, we neglect the $\alpha'^2$ order terms and show that
the Einstein equations are satisfied in this approximation.

In our ansatz for the metric and fluxes, the equations of motion reduce to
\begin{eqnarray}
 E_{(mn)} &=& R_{mn} + 2\nabla_{m}\nabla_{n}\Phi - \frac{1}{4}H_{mpq}{H_{n}}^{pq} \nn
  && \qquad - 2\alpha'
  \left(
   \mathrm{tr}(F_{mp}{F_{n}}^{p}) - \mathrm{tr}(R^{(+)}_{mp}{R^{(+)}_{n}}{^{p}})
  \right) = 0, \label{eq:eom1}\\
 E &=& (\nabla\Phi)^{2} - \frac{1}{2}\dal\Phi
  - \frac{1}{4\cdot 3!}H_{mnp}H^{mnp} + \frac{1}{4}h^{2} \nn
  && \qquad - \frac{\alpha'}{4}
  \left(
   \mathrm{tr}(F_{mn}F^{mn}) - \mathrm{tr}(R^{(+)}_{mn}R^{(+)mn})
  \right) = 0, \label{eq:eom2}\\
 E_{[mn]} &=& \nabla_{p}(e^{-2\Phi}{H^{p}}_{mn}) = 0, \label{eq:eom3}\\
 E_{m} &=& \mathcal{D}^{(-)}_{n}(e^{-2\Phi}{F^n}_m) = 0.\label{eq:eom4}
\end{eqnarray}
The Bianchi identities,
including the leading $\alpha'$-correction, on the other hand, become
\begin{eqnarray}
B_{mnpq} &\equiv& 4\nabla_{[m}H_{npq]}
 - 12\alpha'
  \left(
   \mathrm{tr}(R^{(+)}_{[mn}R^{(+)}_{pq]})
   - \mathrm{tr}(F_{[mn}F_{pq]})
  \right) = 0, \label{eq:bianchi1}\\
 B_{mnp} &\equiv& 3\nabla_{[m}F_{np]} = 0\label{eq:bianchi2}.
\end{eqnarray}
Supersymmetric solutions, in general, must satisfy all these equations
in addition to the Killing spinor equations. However, all these equations
are not independent as can be seen below.

From the integrability conditions of the Killing spinor equations
\begin{eqnarray}
 \left[\nabla^{(-)}_{m},\nabla^{(-)}_{n}\right]\ep &=& 0,\\
 \left[\nabla^{(-)}_{m}, \nabla_{n}\Phi
  - \frac{1}{4\cdot 3!}H_{npq}\gamma^{npq} + \frac{i}{2}h
  \right]\ep &=& 0, \\
 \left[\mathcal{D}^{(-)}_m,F_{mn}\gamma^{mn}\right]\epsilon&=&0,\\
 \left[F_{mn}\gamma^{mn},\nabla_{p}\Phi
 - \frac{1}{4\cdot 3!}H_{qrs}\gamma^{qrs} + \frac{i}{2}h
 \right]\ep &=& 0,
\end{eqnarray}
and identities
\begin{eqnarray}
 (\nabla_{m}\Phi - \frac{1}{2\cdot 3!}H_{mnp}\gamma^{mnp} + \frac{i}{2}h)^{2}\ep&=& 0, \\
 (\nabla_{m}\Phi\gamma^{m})(F_{np}\gamma^{np})\ep &=& 0, \\
 (F_{mn}\gamma^{mn})^{2}\ep &=& 0, \\
 (R^{(+)}_{mnpq}\gamma^{mn})(R^{(-)}_{rstu}\gamma^{tu})\ep &=& 0,
\end{eqnarray}
all of which are satisfied by the solutions of the Killing spinor equations,
one can obtain
\begin{eqnarray}
 2e^{2\Phi}E_{m}\gamma^{m}\ep
  + \frac{1}{3}B_{mnp}\gamma^{mnp}\ep &=& 0,\label{eq:integrability1} \\
 (E_{(mn)} + \frac{1}{2}e^{2\Phi}E_{[mn]})\gamma^{n}\ep
  + \frac{1}{12}B_{mnpq}\gamma^{npq}\ep \nn
  + \frac{\alpha'}{2}R^{(+)}_{mn}{}^{pq}B_{pqrs}\gamma^{n}\gamma^{rs}\ep &=& 0,
\label{eq:integrability2} \\
 2E\ep + \frac{1}{4}e^{2\Phi}E_{[mn]}\gamma^{mn}\ep
  + \frac{1}{16\cdot 3}B_{mnpq}\gamma^{mnpq}\ep \nn
  + \frac{\alpha'}{8}R^{(+)}_{mn}{}^{pq}B_{pqrs}\gamma^{mn}\gamma^{rs}\ep &=& 0,
\label{eq:integrability3}
\end{eqnarray}
including the leading  $\alpha'$ corrections.
By imposing the Bianchi identities (\ref{eq:bianchi1})--(\ref{eq:bianchi2}), 
these equations become
\begin{eqnarray}
 E_{m}\gamma^{m}\ep &=& 0, \\
 (E_{(mn)} + \frac{1}{2}e^{2\Phi}E_{[mn]})\gamma^{n}\ep &=& 0, \\
 2E\ep + \frac{1}{4}e^{2\Phi}E_{[mn]}\gamma^{mn}\ep &=& 0.
\end{eqnarray}
However, from a simple calculation, we can prove that they are equivalent to
the equations of motion (\ref{eq:eom1})--(\ref{eq:eom4}),
\begin{equation}
 E = E_{m} = E_{(mn)} = E_{[mn]} = 0,
\end{equation}
up to $\alpha'^{2}$ order.
Therefore, the equations of motion are automatically satisfied if we impose
the Bianchi identities in addition to the Killing spinor equations.

\section{Summary}
\label{sec:summary}

In this paper, we present the $G$-structure classification for
the $AdS_{3}$-type solutions of Killing spinor equations
in heterotic supergravity.
Their solutions automatically satisfy all the equations of motion, if we impose
the Bianchi identities, which include the leading-order $\alpha'$ 
corrections \cite{Bergshoeff:1989de}.
Here, it is important that there is no first-order 
$\alpha'$-correction in the Killing spinor equations.

By choosing a special local Lorentz frame, we first simplify the Killing spinor
equations before solving them explicitly \cite{Gillard:2004xq}.
Then, we solve the Killing spinor equations and classify the solutions by $G$-structures
in the cases that the numbers of Killing spinors in seven-dimensional manifolds $\mathcal{M}_7$
are $N=1,2,3$, and $4$. These $G$-structures are further classified by
their torsions. The torsion classes for $\mathcal{M}_7$ described by the solutions
of the Killing spinor equations are computed.
Finally, we also study the integrability conditions of the Killing spinor equations 
and show that the Killing spinor equations and Bianchi identities imply all the equations 
of motion, which include the leading $\alpha'$-corrections.
Since we know that there is no leading-order $\alpha'$-correction 
in the Killing spinor equations, the leading $\alpha'$-corrections 
of the supersymmetric solutions only come from the correction of the Bianchi 
identities.\footnote{See also Ref.~\citen{Gillard:2005ic}}

For each case of $N=1,2,3$, and $4$, the solutions of the Killing spinor equations have
the following properties.

For $N=1$, $\mathcal{M}_{7}$ admits the $G_{2}$-structure,
defined by $(\sigma,*\sigma)$. Among the general $G_2$-structures
classified by four torsion classes $(\tau_0,\tau_1,\tau_2,\tau_3)$ as in \S \ref{sec:g-1}, 
the $G$-structure obtained by the Killing spinor is the special one with $\tau_2=0$.

For $N=2$, $\mathcal{M}_7$ admits $SU(3)$-structure, defined by $(K,J,\Omega)$. 
We can obtain a Killing vector from $K$, which decomposes, at least locally, 
$\mathcal{M}_7$ into the orbit $\mathbb{R}$ and its orthogonal six-dimensional submanifold 
$\mathcal{M}_{6}$. The remaining $(J,\Omega)$ can also be interpreted as the $SU(3)$-structure of 
the $\mathcal{M}_6$.
The general $SU(3)$-structure is classified by thirteen torsion classes $(V_1,\cdots,V_8)$ 
and $(\mathcal{W}_1,\cdots,\mathcal{W}_5)$, where the latter five can be interpreted as
the torsion classes of $\mathcal{M}_6$. On the other hand, 
the torsion classes of the $SU(3)$-structure constructed from the Killing spinors must be
$V_2=V_4=V_6=V_7=V_8=0$ and $\mathcal{W}_1=\mathcal{W}_2=0$.
In particular, vanishing torsion classes $\mathcal{W}_1$ and $\mathcal{W}_2$
yield that a submanifold $\mathcal{M}_6$ is a complex manifold.

For $N=3$, $\mathcal{M}_{7}$ is characterized by the $SU(2)$-structure defined by $(K^I,J,\Omega)$.
There are three Killing vectors obtained from $K^I$. 
They decompose $\mathcal{M}_7$ into the orbit $\mathbb{R}^3$
and an orthogonal four-dimensional submanifold $\mathcal{M}_4$ with an $SU(2)$-structure $(J,\Omega)$.
The torsion class of the $SU(2)$-structure obtained from the Killing spinors is given in \S \ref{sec:g-3}.

There is no additional independent Killing spinor bilinear form for $N=4$.
The manifold $\mathcal{M}_{7}$ also has the same $SU(2)$-structure as in the $N=3$ case.
However, the Killing spinor equations impose additional conditions on the three-form $H$.
The torsion class of the $SU(2)$-structure is much more restrictive
than the case of $N=3$.

To obtain the interesting supersymmetric classical solutions, 
we must further impose the Bianchi identities including the leading $\alpha'$-corrections.
We hope to discuss this issue elsewhere.

\section*{Acknowledgements}

The authors would like to thank Tetsuji Kimura for helpful discussions.
They would also like to acknowledge Stefan Ivanov for 
bringing Refs.~\citen{Friedrich:2001yp} and \citen{Friedrich:2001nh} to their attention.
This work is supported in part by the Grant-in-Aid for the Global COE Program 
\lq\lq The Next Generation of Physics, Spun from Universality and Emergence'', 
while the work of H.K. is supported in part by the Grant-in-Aid for Scientific Research (No. 19540284),
both from the Ministry of Education, Culture, Sports, Science and Technology (MEXT) of Japan.

\appendix

%

\section{Conventions for Gamma Matrices}\label{sec:conventions}

In this paper, we adopt a special representation 
for the ten-dimensional gamma matrices $\Gamma^{\hat{M}}$
$(\hat{M}=0,1,\cdots,9)$ as
\begin{eqnarray}
 \Gamma^{\hat{\mu}} &=& 
	\sigma^{1} \otimes \tilde{\gamma}^{\hat{a}} \otimes \mathbf{1},\\
 \Gamma^{\hat{m} + 2} &=& 
	\sigma^{2} \otimes \mathbf{1} \otimes \gamma^{\hat{m}},
\end{eqnarray}
where $\tilde{\gamma}^{\hat{\mu}}$ $(\hat{\mu}=0,1,2)$ and $\gamma^{\hat{m}}$
$(\hat{m}=1,\cdots,7)$ are three- and seven-dimensional gamma matrices, respectively, 
defined by
\begin{equation}
 \tilde{\gamma}^{\hat{0}} = i\sigma^{2},\quad
 \tilde{\gamma}^{\hat{1}} = \sigma^{1}, \quad
 \tilde{\gamma}^{\hat{2}} = \sigma^{3},
\end{equation}
and
\begin{eqnarray}
 \gamma^{\hat{1}} &=& 
	\sigma^{1} \otimes \mathbf{1} \otimes \mathbf{1},\\
 \gamma^{\hat{2}} &=& 
	\sigma^{2} \otimes \mathbf{1} \otimes \mathbf{1},\\
 \gamma^{\hat{3}} &=& 
	\sigma^{3} \otimes \sigma^{1} \otimes \mathbf{1},\\
 \gamma^{\hat{4}} &=& 
	\sigma^{3} \otimes \sigma^{2} \otimes \mathbf{1},\\
 \gamma^{\hat{5}} &=& 
	\sigma^{3} \otimes \sigma^{3} \otimes \sigma^{1}, \\
 \gamma^{\hat{6}} &=& 
	\sigma^{3} \otimes \sigma^{3} \otimes \sigma^{2}, \\
 \gamma^{\hat{7}} &=& 
	-\sigma^{3} \otimes \sigma^{3} \otimes \sigma^{3}, \\
 &=& -i\gamma^{\hat{1}}\gamma^{\hat{2}}\cdots\gamma^{\hat{6}}.\nonumber
\end{eqnarray}
Here, the hatted symbols denote the indices of local Lorentz space.
In this convention, the ten-dimensional chirality operator
$\Gamma^{\natural}$ is
\begin{eqnarray}
 \Gamma^{\natural} &\equiv& \Gamma^{\hat{0}}\Gamma^{\hat{1}}\cdots\Gamma^{\hat{9}} \nonumber \\
 &=& \sigma^{3} \otimes \mathbf{1} \otimes \mathbf{1}.
\end{eqnarray}
The charge conjugation matrices of three and seven dimensions are given as
\begin{eqnarray}
C_{10}&=&1\otimes C_3\otimes C_7,\\
C_3&=&\tilde{\gamma}^{\hat{0}}=i\sigma^2,\\
C_7&=&i\gamma^{\hat{2}}\gamma^{\hat{4}}\gamma^{\hat{6}}
=-\sigma^2\otimes\sigma^1\otimes\sigma^2,\label{eq:ccseven} 
\end{eqnarray}
respectively.

\section{Decompositions of Forms by $G$-Structures}\label{sec:decomp}

For the seven-dimensional manifolds described by the solutions of the Killing spinor
equations, it is convenient to decompose the seven-dimensional local Lorentz vector 
$A_{\hat{m}}$ to $(A_i,A_{\bar{i}},A_7)$, where
\begin{eqnarray}
A_i&=&\frac{1}{\sqrt{2}}\left(A_{\hat{2i-1}}+iA_{\hat{2i}}\right),
\qquad (i=1,2,3)\label{eq:complex}\\
A_{\bar{i}}&=&(A_i)^\dag.
\end{eqnarray}
One can easily extend it to general tensors.
The two-form $F_{\hat{m}\hat{n}}$ is, for example, decomposed as
$(F_{ij},F_{i\bar{j}},F_{\bar{i}\bar{j}},F_{i7},F_{\bar{i}7})$, where
$F_{\bar{i}\bar{j}}=(F_{ij})^\dag$, $F_{\bar{i}7}=(F_{i7})^\dag$
and $F_{\bar{i}j}=(F_{i\bar{j}})^\dag$. 
Each component is defined similarly to (\ref{eq:complex}) as
\begin{eqnarray}
F_{ij}&=&\frac{1}{2}\left(F_{\hat{(2i-1)}\hat{(2j-1)}}+iF_{\hat{2i}\hat{(2j-1)}}
+i\left(F_{\hat{(2i-1)}\hat{2j}}+iF_{\hat{2i}\hat{2j}}\right)\right),\\
F_{i\bar{j}}&=&\frac{1}{2}\left(F_{\hat{(2i-1)}\hat{(2j-1)}}+iF_{\hat{2i}\hat{(2j-1)}}
-i\left(F_{\hat{(2i-1)}\hat{2j}}+iF_{\hat{2i}\hat{2j}}\right)\right),\\
F_{i7}&=&\frac{1}{\sqrt{2}}\left(F_{\hat{2i-1}\hat{7}}+iF_{\hat{2i}\hat{7}}\right).
\end{eqnarray}

\subsection{$G_2$ structure}\label{sec:g2}

The seven-dimensional manifold with a Killing spinor admits a $G_2$ structure with a fundamental 
three-form $\sigma$. The non-trivial fluxes can be decomposed into irreducible representations 
of $G_2$, using $\sigma$ and it's Hodge dual $*\sigma$.

The two-form flux $F$, having twenty-one components, is decomposed 
into seven- and fourteen-dimensional representations of $G_2$ as
\begin{equation}
 F = F^{(7)}\lrcorner\sigma + F^{(14)},
\end{equation}
where
\begin{eqnarray}
 F^{(7)} &\equiv& \frac{1}{3!}F_{mn}{\sigma^{mn}}_{p}dy^p.
\end{eqnarray}

Similarly, the three-form flux $H$, having thirty-five components, is decomposed 
into singlet, seven- and twenty-seven-dimensional representations as
\begin{equation}
 H = H^{(1)}\sigma + H^{(7)}\lrcorner\ast\sigma + H^{(27)},
\end{equation}
where
\begin{eqnarray}
 H^{(1)} &\equiv& \frac{1}{7\cdot 3!}H_{mnp}\sigma^{mnp},\\
 H^{(7)} &\equiv& \frac{1}{4\cdot 3!}H_{mnp}{(\ast\sigma)_q}^{mnp}dy^q.
\end{eqnarray}

For the spin connection, $(\omega_m^{(-)})_{nl}
=(\omega_m^{(-)})_{\hat{n}\hat{l}}e^{\hat{n}}_ne^{\hat{l}}_l$, 
it is convenient to use such a reducible decomposition,
that is, only the antisymmetric group indices $(n,l)$ are decomposed 
into the same form with the two-form:
\begin{equation}
(\omega_m^{(-)})_{nl}=(\omega_m^{(-)})^{(7)p}\sigma_{nlp}+(\omega_m^{(-)})^{(14)}_{nl},
\end{equation}
where
\begin{eqnarray}
 (\omega_m^{(-)})^{(7)p} &=& \frac{1}{3!}(\omega_m^{(-)})_{nl}{\sigma^{nlp}}.
\end{eqnarray}

\subsection{$SU(3)$ structure}\label{sec:su3}

The seven-dimensional manifold with two Killing spinors admits an $SU(3)$ structure defined by $(K,J,\Omega)$. 
The metric can be written as
\begin{equation}
 \d s^{2} = g_{mn}\d y^{m}\d y^{n} + 
(\d y^{7} + w)^{2},
\end{equation}
where $g_{mn}$ is a metric of six-dimensional submanifold
$\mathcal{M}_6$ satisfying ${\mathcal{P}_m}^p{\mathcal{P}_n}^qg_{pq}=g_{mn}$, where
${\mathcal{P}_m}^n={\delta_m}^n-K_mK^n$ is the projection operator onto $\mathcal{M}_6$.
Two- and three-forms $(J,\Omega)$, which also satisfy
${\mathcal{P}_m}^p{\mathcal{P}_n}^qJ_{pq}=J_{mn}$
and ${\mathcal{P}_m}^p{\mathcal{P}_n}^q{\mathcal{P}_l}^r\Omega_{pqr}=\Omega_{mnl}$,
can also be interpreted as an $SU(3)$-structure on $\mathcal{M}_6$. 
From the two-form $J$, we can define an almost complex structure
${J_m}^n$ on $\mathcal{M}_6$. The three-form $\Omega$ is holomorphic
in the sense that it satisfies
${\mathcal{P}_m}^{(-)p}{\mathcal{P}_n}^{(-)q}{\mathcal{P}_l}^{(-)r}\Omega_{pqr}=\Omega_{mnl}$,
where the operator ${\mathcal{P}_m}^{(-)n}$ (${\mathcal{P}_m}^{(+)n}$) defined by
\begin{equation}
{\mathcal{P}_m}^{(\pm)n} \equiv \frac{1}{2}\left({\mathcal{P}_m}^n \pm i{J_{m}}^{n}\right)
\end{equation}
projects a form onto its holomorphic (anti-holomorphic) component.

Then a two-form $F$ $(21)$ on $\mathcal{M}_{7}$ can be decomposed to
a two-form $F_{[6]}$ $(15)$ and a one-form $\tilde{F}_{[6]}$ $(6)$ on $\mathcal{M}_6$,
where the number in the parenthesis is the number of components, as
\begin{equation}
 F = F_{[6]} + \tilde{F}_{[6]}\wedge K,
\end{equation}
where
\begin{eqnarray}
 F_{[6]mn}&=&{\mathcal{P}_m}^p{\mathcal{P}_n}^qF_{pq},\\
 \tilde{F}_{[6]m}&=&{\mathcal{P}_m}^pK^qF_{pq}.
\end{eqnarray}
These forms $F_{[6]}$ $(15)$ and $\tilde{F}_{[6]}$ $(6)$ on $\mathcal{M}_6$ 
are further decomposed, using the $SU(3)$-structure $(J,\Omega)$, into $1+3+\bar{3}+8$-
dimensional representations as
\begin{eqnarray}
 F_{[6]} &=& F^{(1)}J+ 
F^{(3)}\lrcorner\bar{\Omega} + F^{(\bar{3})}\lrcorner\Omega
  + F^{(8)}, \\
 F^{(1)} &\equiv& \frac{1}{6}F_{mn}J^{mn},\\
 F^{(3)} &\equiv& \frac{1}{8\cdot 2!}F_{mn}{\Omega^{mn}}_{p}dy^p,\\
 F^{(\bar{3})} &\equiv& \frac{1}{8\cdot 2!}F_{mn}{\bar{\Omega}^{mn}}_{p}dy^p,
\end{eqnarray}
and $3+\bar{3}$-dimensional representations as
\begin{eqnarray}
 \tilde{F}_{[6]} &=& \tilde{F}^{(3)} + \tilde{F}^{(\bar{3})}, \\
 \tilde{F}^{(3)} &\equiv& {\mathcal{P}_n}^{(-)m}\tilde{F}_{[6]m}dy^n,\\
 \tilde{F}^{(\bar{3})} &\equiv& {\mathcal{P}_n}^{(+)m}\tilde{F}_{[6]m}dy^n.
\end{eqnarray}

A three-form $H$ $(35)$ can also be decomposed to a three-form $H_{[6]}$ $(20)$
and a two-form $\tilde{H}_{[6]}$ $(15)$ on $\mathcal{M}_6$ as
\begin{equation}
 H = H_{[6]} + \tilde{H}_{[6]}\wedge K,
\end{equation}
where
\begin{eqnarray}
H_{[6]mnl}&=&{\mathcal{P}_m}^p{\mathcal{P}_n}^q{\mathcal{P}_l}^rH_{pqr},\\
\tilde{H}_{[6]mn}&=&{\mathcal{P}_m}^p{\mathcal{P}_n}^qK^rH_{pqr}. 
\end{eqnarray}
These forms on $\mathcal{M}_6$ are further decomposed into
$20=1+\bar{1}+3+\bar{3}+6+\bar{6}$ as
\begin{eqnarray}
 H_{[6]} &=& H^{(1)}\bar{\Omega} +  H^{(\bar{1})}\Omega
  + (H^{(3)} + H^{(\bar{3})})\wedge J + H^{(6)} + H^{(\bar{6})}, \\
 H^{(1)} &\equiv& \frac{1}{8\cdot 3!}H_{[6]mnp}\Omega^{mnp},\\
 H^{(\bar{1})} &\equiv& \frac{1}{8\cdot 3!}H_{[6]mnp}\bar{\Omega}^{mnp},\\
 H^{(3)} &\equiv& \frac{1}{4}H_{[6]mnp}J^{np}{\mathcal{P}_n}^{(-)m}dy^n,\\
 H^{(\bar{3})} &\equiv& \frac{1}{4}H_{[6]mnp}J^{np}{\mathcal{P}_n}^{(+)m}dy^n,\\
 H^{(6)} &\equiv&
	\frac{1}{2!}(H_{[6]} - H^{(3)}\wedge J)_{mnp}
	{\mathcal{P}_q}^{(-)m}{\mathcal{P}_s}^{(-)n}{\mathcal{P}_t}^{(+)p}
            dy^q\wedge dy^s\wedge dy^t,\\
 H^{(\bar{6})} &\equiv&	
	\frac{1}{2!}(H_{[6]} - H^{(\bar{3})}\wedge J)_{mnp}
	{\mathcal{P}_q}^{(+)m} {\mathcal{P}_s}^{(+)n} {\mathcal{P}_t}^{(-)p}
            dy^q\wedge dy^s\wedge dy^t,
\end{eqnarray}
and $15=1+3+\bar{3}+8$ as
\begin{eqnarray}
 \tilde{H}_{[6]} &=& \tilde{H}^{(1)}J + \tilde{H}^{(3)}\lrcorner\bar{\Omega} + 
\tilde{H}^{(\bar{3})}\lrcorner\Omega +  \tilde{H}^{(8)}, \\
 \tilde{H}^{(1)} &\equiv& \frac{1}{6}\tilde{H}_{[6]mn}J^{mn},\\
 \tilde{H}^{(3)} &\equiv& \frac{1}{8\cdot 2!}\tilde{H}_{[6]mn}{\Omega^{mn}}_{q}dx^{q},\\
 \tilde{H}^{(\bar{3})} &\equiv& \frac{1}{8\cdot 2!}\tilde{H}_{[6]mn}{\bar{\Omega}^{mn}}_{q}dx^{q}.
\end{eqnarray}

Similar to the two form $F$, the spin connection can first be decomposed as
\begin{eqnarray}
(\omega_m^{(-)})_{nl}&=&(\omega_m^{(-)})_{[6]nl}+2(\tilde{\omega}_m^{(-)})_{[6][n}K_{l]},
\end{eqnarray}
where
\begin{eqnarray}
 (\omega_m^{(-)})_{[6]nl}&=&{\mathcal{P}_n}^p{\mathcal{P}_l}^q(\omega_m^{(-)})_{pq},\\
 (\tilde{\omega}_m^{(-)})_{[6]n}&=&{\mathcal{P}_n}^pK^q(\omega_m^{(-)})_{pq}.
\end{eqnarray}
These are further decomposed as
\begin{eqnarray}
(\omega_m^{(-)})_{[6]nl}&=&(\omega_m^{(-)})^{(3)p}\bar{\Omega}_{pnl}
+(\omega_m^{(-)})^{(\bar{3})p}\Omega_{pnl}
+(\omega_m^{(-)})^{(1)}J_{nl}+(\omega_m^{(-)})^{(8)}_{nl},\\
(\omega_m^{(-)})^{(3)p}&=&\frac{1}{8\cdot 2!}(\omega_m^{(-)})_{[6]nl}\Omega^{nlp},\\
(\omega_m^{(-)})^{(\bar{3})p}&=&\frac{1}{8\cdot 2!}(\omega_m^{(-)})_{[6]nl}\bar{\Omega}^{nlp},\\
(\omega_m^{(-)})^{(1)}&=&\frac{1}{6}(\omega_m^{(-)})_{[6]nl}J^{nl}, 
\end{eqnarray}
and
\begin{eqnarray}
(\tilde{\omega}_m^{(-)})_{[6]n}&=&(\tilde{\omega}_m^{(-)})^{(3)}_n
+(\tilde{\omega}_m^{(-)})^{(\bar{3})}_n,\\
(\tilde{\omega}_m^{(-)})^{(3)}_n&=&(\tilde{\omega}_m^{(-)})_{[6]p}{\mathcal{P}_n}^{(-)p},\\
(\tilde{\omega}_m^{(-)})^{(\bar{3})}_n&=&(\tilde{\omega}_m^{(-)})_{[6]p}{\mathcal{P}_n}^{(+)p}. 
\end{eqnarray}

\subsection{$SU(2)$ structure}\label{sec:su2}

The seven-dimensional manifold with three or four Killing spinors admits an $SU(2)$ structure,
defined by three one-forms, a two-form, and a three-form $(K^{(I)},J,\Omega)$.
The metric can be written as
\begin{equation}
 \d s^{2} = g_{mn}\d y^{m}\d y^{n}+(\d y^{5} + w_{(2)})^{2} + (\d y^{6} + w_{(3)})^{2}
	+ (\d y^{7} + w_{(1)})^{2},
\end{equation}
where $g_{mn}$ is a metric of four-dimensional submanifold $\mathcal{M}_4$
satisfying ${\mathcal{P}_m}^p{\mathcal{P}_n}^qg_{pq}=g_{mn}$,
where ${\mathcal{P}_m}^n={\delta_m}^n-\sum_IK^{(I)}_mK^{(I)n}$ is the projection
operator onto $\mathcal{M}_4$. Two- and three-forms $(J,\Omega)$, 
which also satisfy  ${\mathcal{P}_m}^p{\mathcal{P}_n}^qJ_{pq}=J_{mn}$
and ${\mathcal{P}_m}^p{\mathcal{P}_n}^q{\mathcal{P}_l}^r\Omega_{pqr}=\Omega_{mnl}$,
can also be interpreted as an $SU(2)$-structure on $\mathcal{M}_4$. 
From the two-form $J$, we can define an almost complex structure ${J_m}^n$ on
$\mathcal{M}_4$. The three-form $\Omega$ is holomorphic
in the sense that it satisfies
${\mathcal{P}_m}^{(-)p}{\mathcal{P}_n}^{(-)q}{\mathcal{P}_l}^{(-)r}\Omega_{pqr}=\Omega_{mnl}$,
where the operator ${\mathcal{P}_m}^{(-)n}$ (${\mathcal{P}_m}^{(+)n}$) defined by
\begin{equation}
{\mathcal{P}_m}^{(\pm)n} \equiv \frac{1}{2}\left({\mathcal{P}_m}^n \pm i{J_{m}}^{n}\right)
\end{equation}
projects a form onto its holomorphic (antiholomorphic) component.

Then a two-form $F$ $(21)$ on $\mathcal{M}_7$ can be decomposed to a two-form $F_{[4]}$ $(6)$,
three one-form $F_{[4]I}$ $(4)$, and three zero-form $F_{IJ}$ $(1)$  on $\mathcal{M}_4$ as
\begin{equation}
 F = F_{[4]} 
	+ F_{[4]I}\wedge K^{I} 
	+ \frac{1}{2!}F_{IJ}\wedge K^{I}\wedge K^{J},
\end{equation}
where
\begin{eqnarray}
(F_{[4]})_{mn}&=&{\mathcal{P}_m}^p{\mathcal{P}_n}^qF_{pq},\\
(F_{[4]I})_m&=&{\mathcal{P}_m}^pK^q_{(I)}F_{pq},\\
(F_{IJ})&=&K^p_{(I)}K^q_{(J)}F_{pq}. 
\end{eqnarray}
These forms $F_{[4]}$ $(6)$  and $F_{[4]I}$ $(4)$ on $\mathcal{M}_4$
are further decomposed, using $SU(2)$ structure $(J,\Omega)$, into $6=1+\bar{1}+1'+3$ as
\begin{eqnarray}
F_{[4]}&=&F^{(1)}\bar{\Omega}+F^{(\bar{1})}\Omega+F^{(1')}J+F^{(3)},\\
F^{(1)}&=&\frac{1}{8}(F_{[4]})_{mn}\Omega^{mn},\\ 
F^{(\bar{1})}&=&\frac{1}{8}(F_{[4]})_{mn}\bar{\Omega}^{mn},\\
F^{(1')}&=&\frac{1}{4}(F_{[4]})_{mn}J^{mn},\\ 
\end{eqnarray}
and $4=2+\bar{2}$ as
\begin{eqnarray}
F_{[4]I}&=&F^{(2)}_I+F^{(\bar{2})}_I,\\ 
F^{(2)}_I&=&{\mathcal{P}_m}^{(-)p}F_{[4]p}dy^m,\\
F^{(\bar{2})}_I&=&{\mathcal{P}_m}^{(+)p}F_{[4]p}dy^m.
\end{eqnarray}

A three-form $H$ $(35)$ can also be decomposed to a three-form $H_{[4]}$ $(4)$,
three two-form $H_{[4]I}$ $(6)$, three one-form $H_{[4]IJ}$ $(4)$, and a zero-form
$H_{IJK}$ $(1)$ on $\mathcal{M}_4$ as
\begin{equation}
 H = H_{[4]} 
	+ H_{[4]I}\wedge K^{I} 
	+ \frac{1}{2!}H_{[4]IJ}\wedge K^{I}\wedge K^{J}
	+ \frac{1}{3!}\tilde{H}_{IJK}\wedge K^{I}\wedge K^{J}\wedge K^{K},
\end{equation}
where
\begin{eqnarray}
(H_{[4]})_{mnl}&=&{\mathcal{P}_m}^p{\mathcal{P}_n}^q{\mathcal{P}_l}^rH_{pqr},\\
(H_{[4]I})_{mn}&=&{\mathcal{P}_m}^p{\mathcal{P}_n}^qK^r_{(I)}H_{pqr},\\
(H_{[4]IJ})_m&=&{\mathcal{P}_m}^pK^q_{(I)}K^r_{(J)}H_{pqr},\\
(\tilde{H}_{[4]IJK})&=&K^p_{(I)}K^q_{(J)}K^r_{(K)}H_{pqr}.
\end{eqnarray}
The three-form $H_{[4]}$ $(4)$ on $\mathcal{M}_4$ is further decomposed
into $4=2+\bar{2}$ as
\begin{eqnarray}
 H_{[4]} &=& H^{(2)}\wedge J + H^{(\bar{2})}\wedge J, \\
 H^{(2)} &\equiv& \frac{1}{2}H_{[4]pqr}J^{qr}{\mathcal{P}_m}^{(-)p}dy^m, \\
 H^{(\bar{2})} &\equiv& \frac{1}{2}H_{[4]pqr}J^{qr}{\mathcal{P}_m}^{(+)p}dy^m.
\end{eqnarray}
Similarly, the two-form $H_{[4]I}$ $(6)$ and one-form $H_{[4]IJ}$ $(4)$ are decomposed
into $6=1+\bar{1}+1'+3$ as
\begin{eqnarray}
 H_{[4]I} &=& H^{(1)}_{I}\bar{\Omega} + H_{I}^{(\bar{1})}\Omega
  + H^{(1^{\prime})}_{I}J + H^{(3)}_{I}, \\
 H^{(1)}_{I} &\equiv& \frac{1}{8}(H_{[4]I})_{mn}\Omega^{mn},\\
 H^{(\bar{1})}_{I} &\equiv& \frac{1}{8}(H_{[4]I})_{mn}\bar{\Omega}^{mn},\\
 H^{(1^{\prime})}_{I} &\equiv& \frac{1}{4}(H_{[4]I})_{mnp}J^{mn},
\end{eqnarray}
and $4=2+\bar{2}$ as
\begin{eqnarray}
 H_{[4]IJ} &=& H^{(2)}_{IJ} + H^{(\bar{2})}_{IJ},\\
 H^{(2)}_{IJ} &\equiv& (H_{[4]IJ})_{p}{\mathcal{P}_m}^{(-)p}dy^m,\\
 H^{(\bar{2})}_{IJ} &\equiv& (H_{IJ})_{p}{\mathcal{P}_m}^{(+)p}dy^m.
\end{eqnarray}
We can also define $\tilde{H}_{IJK}=\epsilon_{IJK}\tilde{H}^{(1)}$.

The spin connection can also be decomposed as
\begin{eqnarray}
(\omega_m^{(-)})_{nl}&=&(\omega_m^{(-)})_{[4]nl}+2(\omega_m^{(-)})_{[4]I[n}K_{l]}^I+
(\omega_m^{(-)})_{IJ}K_{[n}^{(I)}K_{l]}^{(J)},\\ 
\end{eqnarray}
where
\begin{eqnarray}
(\omega_m^{(-)})_{[4]nl}&=&{\mathcal{P}_n}^p{\mathcal{P}_l}^q(\omega_m^{(-)})_{pq},\\
(\omega_m^{(-)})_{[4]In}&=&{\mathcal{P}_n}^pK_{(I)}^q(\omega_m^{(-)})_{pq},\\
(\omega_m^{(-)})_{IJ}&=& K_{(I)}^pK_{(J)}^q(\omega_m^{(-)})_{pq}.
\end{eqnarray}
Each component is further decomposed as
\begin{eqnarray}
 (\omega_m^{(-)})_{[4]nl}&=&(\omega_m^{(-)})^{(1)}\bar{\Omega}_{nl}
+(\omega_m^{(-)})^{(\bar{1})}\Omega_{nl}
+(\omega_m^{(-)})^{(1')}J_{nl}+(\omega_m^{(-)})^{(3)}_{nl},\\
(\omega_m^{(-)})^{(1)}&=&\frac{1}{8}(\omega_m^{(-)})_{[4]nl}\Omega^{nl},\\
(\omega_m^{(-)})^{(\bar{1})}&=&\frac{1}{8}(\omega_m^{(-)})_{[4]nl}\bar{\Omega}^{nl},\\
(\omega_m^{(-)})^{(1')}&=&\frac{1}{4}(\omega_m^{(-)})_{[4]nl}J^{nl},
\end{eqnarray}
and
\begin{eqnarray}
(\omega_m^{(-)})_{[4]In}&=&(\omega_m^{(-)})^{(2)}_{In}+(\omega_m^{(-)})^{(\bar{2})}_{In}, \\
(\omega_m^{(-)})^{(2)}_{In}&=&{\mathcal{P}_n}^{(-)p}(\omega_m^{(-)})_{[4]Ip},\\
(\omega_m^{(-)})^{(\bar{2})}_{In}&=&{\mathcal{P}_n}^{(+)p}(\omega_m^{(-)})_{[4]Ip}.
\end{eqnarray}


\providecommand{\href}[2]{#2}\begingroup\raggedright
\end{document}